\documentclass[a4paper,twoside,12pt]{article}
\input{obsjkt.sty}

\newcommand{\Msun}{\ensuremath{~{\rm M}_\odot}}                   
\newcommand{\Rsun}{\ensuremath{~{\rm R}_\odot}}                   
\newcommand{\rhosun}{\ensuremath{~\rho_\odot}}                    
\newcommand{\Teff}{\ensuremath{T_{\rm eff}}}                      
\newcommand{\EBV}{\ensuremath{E(B\!-\!V)}}                        
\newcommand{\degr}{\ensuremath{^\circ}}                           
\renewcommand{\kms}{~km~s$^{-1}$}                                 
\renewcommand{\cd}{~d$^{-1}$}                                     
\newcommand{\etal}{\textit{et al.}}                               
\newcommand{\gaia}{\textit{Gaia}}                                 
\newcommand{\targ}{MU~Cas}
\newcommand{\targfull}{MU~Cassiopeiae}

\newcommand{\Msunnom}{\hbox{$\mathcal{M}^{\rm N}_\odot$}}
\newcommand{\Rsunnom}{\hbox{$\mathcal{R}^{\rm N}_\odot$}}
\newcommand{\Lsunnom}{\hbox{$\mathcal{L}^{\rm N}_\odot$}}

\usepackage{rotating}

\begin{document} 

\OBSheader{Rediscussion of eclipsing binaries: \targ}{J.\ Southworth}{2025 February}

\OBStitle{Rediscussion of eclipsing binaries. Paper XXII. \\ The B-type system MU Cassiopeiae}

\OBSauth{John Southworth}

\OBSinstone{Astrophysics Group, Keele University, Staffordshire, ST5 5BG, UK}


\OBSabstract{\targ\ is a detached eclipsing binary containing two B5~V stars in an orbit of period 9.653~d and eccentricity 0.192, which has been observed in seven sectors using the Transiting Exoplanet Survey Satellite (TESS). We use these new light curves together with published spectroscopic results to measure the physical properties of the component stars, finding masses of $4.67 \pm 0.09$\Msun\ and $4.59 \pm 0.08$\Msun, and radii of $4.12 \pm 0.04$\Rsun\ and $3.65 \pm 0.05$\Rsun. These values agree with previous results save for a change in which of the two stars is designated the primary component. The measured distance to the system, $1814 \pm 37$~pc, is 1.8$\sigma$ shorter than the distance from the \gaia\ DR3 parallax. A detailed spectroscopic analysis of the system is needed to obtain improved temperature and radial velocity measurements for the component stars; a precise spectroscopic light ratio is also required for better measurement of the stellar radii. \targ\ matches the predictions of theoretical stellar evolutionary models for a solar chemical composition and an age of $87 \pm 5$~Myr. No evidence for pulsations was found in the light curves.}


\section*{Introduction}

The study of detached eclipsing binaries (dEBs) allows the direct and high-precision measurement of the masses and radii of stars \cite{Andersen91aarv,Torres++10aarv}, which can be used to confirm and improve the predictions of theoretical models of stellar evolution \cite{Pols+97mn,LastennetVallsgabaud02aa,HiglWeiss17aa}. The recent plethora of space-based telescopes has revolutionised this work \cite{Me21univ} by providing light curves of previously unattainable quality for a large number of dEBs. In the current series of papers \cite{Me20obs} we are using this opportunity to improve and update measurements of known dEBs to increase the number with mass and radius measurements to 2\% precision and accuracy \cite{Me15aspc}.

In this work we study the system \targfull\footnote{Note that entering ``mu cas'' into databases such as \emph{Simbad} returns information for the bright star $\mu$~Cas. The results for the eclipsing binary can sometimes be obtained by searching for ``MU Cas'' (note capitalisation), but in other cases ``V* MU Cas'' or alternative designations such as ``HIP 1263'' are more reliable.} an EB containing two B5~V stars on an eccentric orbit with a period of 9.653~d (Table~\ref{tab:info}). Our analysis relies on new high-quality space-based photometry and on spectroscopic results available in the literature. Our decision to study this object was partly motivated by the recent acquisition of extensive light curves using the TESS mission, and partly by the possibility of including it in an unrelated project (in preparation).

The variability of \targ\ was discovered by Hoffmeiester \cite{Hoffmeister49ane}, and subsequent work has been summarised by Lacy, Claret \& Sabby \cite{Lacy++04aj2} (hereafter LCS04). LCS04 were the first to determine the orbital period of the system correctly, and also measured the properties of the component stars from extensive $V$-band light curves and a set of radial velocities (RVs) from high-resolution spectra.

Claret \etal \cite{Claret+21aa} measured the apsidal motion of the system, which is slow, and did not use it as it lacked sufficient precision for their analysis. Aside from this work, \targ\ has been mentioned in a multitude of catalogue papers and lists of observed times of minimum brightness which need not be itemised here.

LCS04 deduced photometric spectral types of B5~V for both components of \targ\ based on $UBV$ photometry. We define star~A to be the star eclipsed at the primary (deeper) eclipse, and star~B to be its companion. By this definition, star~A turns out to be the larger and more massive of the two, but has evolved to a cooler effective temperature (\Teff).



\begin{table}[t]
\caption{\em Basic information on \targfull. The $BV$ magnitudes are each the mean of 122 individual measurements \cite{Hog+00aa} distributed approximately randomly in orbital phase, and agree well with the out-of-eclipse values from Lacy \cite{Lacy92aj}. The $JHK_s$ magnitudes are from 2MASS \cite{Cutri+03book} and were obtained at orbital phase 0.268. \label{tab:info}}
\centering
\begin{tabular}{lll}
{\em Property}                            & {\em Value}                 & {\em Reference}                      \\[3pt]
Right ascension (J2000)                   & 00 15 51.56                 & \citenum{Gaia23aa}                   \\
Declination (J2000)                       & $+$60 25 53.6               & \citenum{Gaia23aa}                   \\
\textit{Gaia} DR3 designation             & 429158427924463872          & \citenum{Gaia21aa}                   \\
\textit{Gaia} DR3 parallax                & $0.5133 \pm 0.0191$ mas     & \citenum{Gaia21aa}                   \\          
TESS\ Input Catalog designation           & TIC 83905462                & \citenum{Stassun+19aj}               \\
$B$ magnitude                             & $11.12 \pm 0.05$            & \citenum{Hog+00aa}                   \\          
$V$ magnitude                             & $10.80 \pm 0.06$            & \citenum{Hog+00aa}                   \\          
$J$ magnitude                             & $10.127 \pm 0.022$          & \citenum{Cutri+03book}               \\
$H$ magnitude                             & $10.083 \pm 0.021$          & \citenum{Cutri+03book}               \\
$K_s$ magnitude                           & $10.020 \pm 0.016$          & \citenum{Cutri+03book}               \\
Spectral type                             & B5~V + B5~V                 & \citenum{Lacy++04aj2}                \\[3pt]
\end{tabular}
\end{table}


\section*{Photometric observations}

\begin{figure}[t] \centering \includegraphics[width=\textwidth]{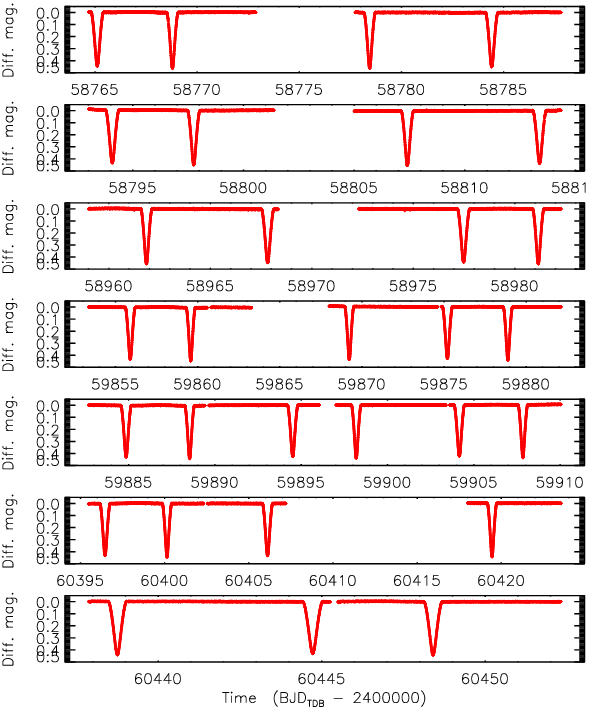} \\
\caption{\label{fig:time} TESS short-cadence SAP photometry of \targ\ from sectors 17, 18, 
24, 57, 58, 77 and 78 (top to bottom panels). The flux measurements have been converted to 
magnitude units then rectified to zero magnitude by subtraction of the median.} \end{figure}


A profusion of photometric data exists for \targ, as it has been observed at 120~s cadence in sectors 17, 18, 24, 57, 58, 77 and 78 by TESS. We downloaded all these data from the NASA Mikulski Archive for Space Telescopes (MAST\footnote{\texttt{https://mast.stsci.edu/portal/Mashup/Clients/Mast/Portal.html}}) using the {\sc lightkurve} package \cite{Lightkurve18} and the quality flag ``hard''. The simple aperture photometry (SAP) light curves from the SPOC data reduction \cite{Jenkins+16spie} pipeline were used, converted into differential magnitude and with the median magnitude subtracted for each sector.

The light curves are shown in Fig.~\ref{fig:time}. Some gaps in coverage exist due to pauses in observation by the spacecraft, or where the quality threshold was not met, and a few instrumental jumps or trends are discernible. There is a total of 105,609 datapoints within these sectors. We trimmed a further set of datapoints where slow instrumental trends were clear, leaving behind 97,571 datapoints.

A query of the \gaia\ DR3 database\footnote{\texttt{https://vizier.cds.unistra.fr/viz-bin/VizieR-3?-source=I/355/gaiadr3}} returns a total of 282 sources within 2~arcmin of \targ, as expected due to the faint limiting magnitude of \gaia\ and the proximity of our target to the galactic plane. \targ\ is the brightest star within this sky region, the second-brightest is distant by 1.33~arcmin and fainter by 1.33~mag in the \gaia\ $G_{\rm RP}$ band, and the next-brightest is at 1.79~arcmin and fainter by 2.62~mag in the same band. As the pixel size and point spread functions of TESS are large, at 21~arcsec and 84~arcsec (90\% encircled energy) respectively, these nearby stars will contribute a small amount of contaminating light to the TESS observations of \targ.



\section*{Light curve analysis}

Our first approach was to isolate the data near eclipse. We extracted the data within 1.1~d of each eclipse midpoint, and renormalised them to zero differential magnitude by fitting a straight line or quadratic function to the data outside eclipse. On inspection of the results it was found that the eclipse depths change slightly between sectors -- the primary eclipses vary from a depth of 0.463~mag (sector 17) to 0.440~mag (sector 58). We attribute this to varying amounts of contaminating light, as the sectors of data were obtained with different spacecraft orientations and pixel masks in the photometry pipeline. We therefore decided to model the sectors individually and combine the results afterwards.

The components of \targ\ are well-separated, so the system is suitable for analysis with the {\sc jktebop}\footnote{\texttt{http://www.astro.keele.ac.uk/jkt/codes/jktebop.html}} code \cite{Me++04mn2,Me13aa} (version 43). We fitted the following parameters for each TESS sector: the fractional radii of the stars ($r_{\rm A}$ and $r_{\rm B}$), expressed as their sum ($r_{\rm A}+r_{\rm B}$) and ratio ($k = {r_{\rm B}}/{r_{\rm A}}$), the central surface brightness ratio ($J$), third light ($L_3$), orbital inclination ($i$), orbital period ($P$), reference time of primary minimum ($T_0$), the orbital eccentricity ($e$) and the argument of periastron ($\omega$) expressed as their Poincar\'e combinations ($e\cos\omega$ and $e\sin\omega$), one limb darkening coefficient (see below), and a linear trend for the out-of-eclipse brightness for each TESS half-sector.

Limb darkening (LD) was included using the power-2 approximation \cite{Hestroffer97aa,Maxted18aa,Me23obs2} and the similarity of the stars allowed the use of the same LD coefficients for both stars. We fitted for the linear coefficient ($c$) but fixed the non-linear coefficient ($\alpha$) to a suitable theoretical value \cite{ClaretSouthworth22aa,ClaretSouthworth23aa}. The strong correlation between $c$ and $\alpha$, and the inclusion of $c$ in the list of fitted parameters, means our results are effectively independent of stellar theory.

\begin{figure}[t] \centering \includegraphics[width=\textwidth]{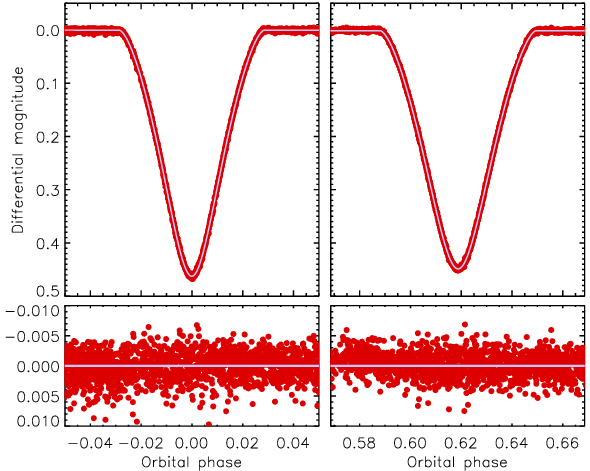} \\
\caption{\label{fig:phase} {\sc jktebop} best fit to the 120-s cadence light curves 
of \targ\ from TESS sector 17. The data are shown as filled red circles and the best 
fit as a light blue solid line. The residuals are shown on an enlarged scale in the 
lower panel.} \end{figure}

The best fit to the data from sector 17 is good, is shown in Fig.~\ref{fig:phase}, and is representative of the results for the other sectors. Once the fits to each of the sectors were established, we ran Monte Carlo and residual-permutation solutions \citep{Me08mn} to obtain errorbars for the measured parameters \citep{Me21obs4}. The immediate outcome of this process was that the results between sectors agree well with each other, but not within the uncertainties. For our final results we therefore provide the unweighted mean for each parameter, with an uncertainties calculated by dividing the standard deviation of the values by the square-root of the number of sectors. These numbers are collected in Table~\ref{tab:jktebop}.

\begin{table} \centering
\caption{\em \label{tab:jktebop} Photometric parameters of \targ\ measured using 
{\sc jktebop}. The value for each parameter is the unweighted mean of the individual 
values per TESS sector, and its uncertainty is the standard deviation of the values 
divided by the square-root of the number of sectors.}
\begin{tabular}{lcc}
{\em Parameter}                           &              {\em Value}            \\[3pt]
{\it Fitted parameters:} \\                                                   
Orbital inclination (\degr)               & $      87.110      \pm  0.033     $ \\
Sum of the fractional radii               & $       0.19395    \pm  0.00027   $ \\
Ratio of the radii                        & $       0.888      \pm  0.016     $ \\
Central surface brightness ratio          & $       1.0178     \pm  0.0008    $ \\
Third light                               & $       0.0334     \pm  0.0061    $ \\
LD coefficient $c$                        & $       0.519      \pm  0.027     $ \\
LD coefficient $\alpha$                   &            0.3811 (fixed)           \\
$e\cos\omega$                             & $       0.18728    \pm  0.00008   $ \\
$e\sin\omega$                             & $       0.04215    \pm  0.00054   $ \\
{\it Derived parameters:} \\                                                   
Fractional radius of star~A               & $       0.10275    \pm  0.00075   $ \\
Fractional radius of star~B               & $       0.09119    \pm  0.00099   $ \\
Light ratio $\ell_{\rm B}/\ell_{\rm A}$   & $       0.804      \pm  0.029     $ \\[3pt]
Orbital eccentricity                      & $       0.19197    \pm  0.00011   $ \\
Argument of periastron ($^\circ$)         & $      12.68       \pm  0.16      $ \\
\end{tabular}
\end{table}

We find that some parameters are determined extremely well; these include the orbital inclination ($\pm$0.03\degr), the sum of the fractional radii (fractional uncertainty of 0.3\%), the central surface brightness ratio (0.2\%) and $e\cos\omega$. However, the ratio of the radii and the light ratio are relatively poorly determined and strongly correlated with other parameters. This effect is common in the analysis of the light curves of dEBs with eclipses that are not total (e.g.\ ref.~\cite{Torres+00aj}) and is due to changes in the ratio of the radii having little effect on the shape of the light curve. 

\begin{figure}[t] \centering \includegraphics[width=\textwidth]{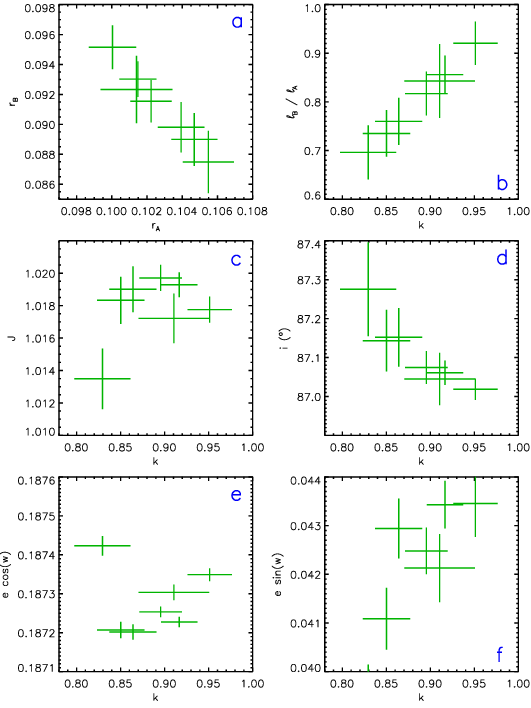} \\
\caption{\label{fig:corr} Correlation plots from the {\sc jktebop} fits to the 
individual TESS sectors. The errorbars in each case are the uncertainties obtained
from Monte Carlo simulations.} \end{figure}

To visualise this we have constructed a set of correlation plots in Fig.~\ref{fig:corr}. Panel (a) shows that the fractional radii of the stars are strongly anti-correlated, as expected when their sum is much better determined than their ratio. Panels (b) and (c) show that the surface brightness ratio is well-constrained (by the relative depths of the eclipses) and thus the uncertainty in the ratio of the radii manifests as a large uncertainty in the light ratio. Panel (d) shows that the correlation is much weaker for the orbital inclination, and panels (e) and (f) that it has no significant effect on the Poincar\'e quantities.

Panel (b) shows that a direct measurement of the light ratio of the two stars, either from a composite spectrum or high-resolution imaging (e.g.\ refs.~\cite{Me++07aa} and \cite{Me21obs2}), could solve this problem by specifying the allowed range of values of the ratio of the radii. To demonstrate this we reran the {\sc jktebop} fit of the sector 78 light curve with the imposition of the spectroscopic light ratio of $0.79 \pm 0.04$ reported by LCS04. The uncertainties in the fractional radii were decreased by roughly a factor of 1.5 with respect to the solution without the light ratio, and application to all sectors results in a tighter clustering of parameter values. Our results are in good agreement with the spectroscopic light ratio, but are independent of it; our tests show that a more precise light ratio than this is needed to better measure the radii of the stars.


\section*{Radial velocity analysis}

\begin{figure}[t] \centering \includegraphics[width=\textwidth]{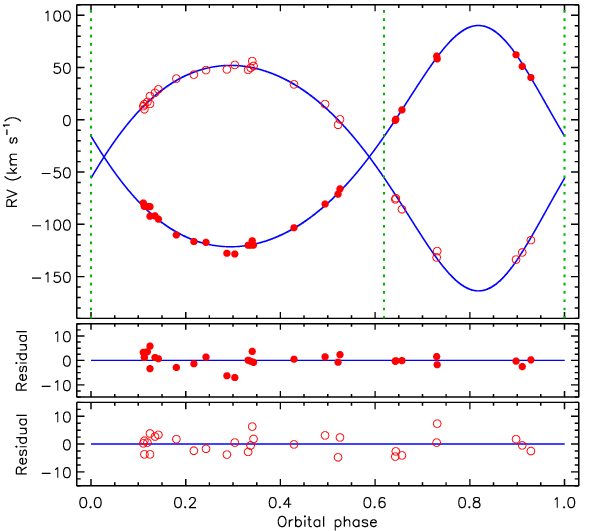} \\
\caption{\label{fig:rv} RVs of \targ\ from LCS04 (filled red circles for star~A and 
open red circles for star~B), compared to the best fit from {\sc jktebop} (solid blue
lines). The times of eclipse are given using vertical green dotted lines. The residuals 
are given in the lower panels separately for the two components.} \end{figure}

The TESS observations allow a more precise photometric model of the system, specifically for the orbital eccentricity and ephemeris. The eccentricity has been precisely measured above, but the ephemeris has not. We therefore fitted the light curve containing each fully-observed eclipse (see above) with {\sc jktebop} to determine a precise ephemeris. We did not include published times of minimum because \targ\ experiences apsidal motion and analysis of that effect is outside the scope of the current work. The resulting ephemeris is
\begin{equation}
  \mbox{Min~I} = {\rm BJD}_{\rm TDB}~ 2459869.229815 (31) + 9.65295307 (29) E
\end{equation}
where $E$ is the number of cycles elapsed since the reference time. 

Armed with this new information, it is worthwhile revisiting the spectroscopic orbit of the system. LCS04 obtained and presented 29 spectroscopic RV measurements for each star, which they fitted together with their photometric observations. We obtained the RVs from their table 2 and performed an independent fit with {\sc jktebop}. We fixed the parameters of the system except for the argument of periastron (to allow for possible apsidal motion), and the velocity amplitudes ($K_{\rm A}$ and $K_{\rm B}$) and systemic velocities ($V_{\rm\gamma,A}$ and $V_{\rm\gamma,B}$) of the stars. We did not force $V_{\rm\gamma,A}$ to equal $V_{\rm\gamma,B}$ but their fitted values agree well. We also allowed for a phase offset versus the orbital ephemeris above, to account for shifts due to apsidal motion or time conversion errors. The data were not provided with errorbars so we adopted a single uncertainty for all RVs per star and adjusted it to force a reduced $\chi^2$ of unity for each star.


The best fit to the LCS04 RVs is shown in Fig.~\ref{fig:rv} and is practically identical to that presented in fig.~1 of LCS04. We found $K_{\rm A} = 105.83 \pm 0.85$\kms, $K_{\rm B} = 107.86 \pm 0.95$\kms, $V_{\rm\gamma,A} = -35.57 \pm 0.55$\kms\ and $V_{\rm\gamma,B} = -35.49 \pm 0.66$\kms. The argument of periastron ($\omega = 10.8 \pm 1.4$\degr) agrees with the photometric value in Table~\ref{tab:jktebop}, and the phase offset [$\Delta\phi = (8 \pm 9) \times 10^{-5}$] is consistent with zero. The errorbars quoted here were obtained from Monte Carlo simulations \cite{Me21obs5}.


\section*{Physical properties and distance to \targ}

\begin{table} \centering
\caption{\em Physical properties of \targ\ defined using the nominal solar units 
given by IAU 2015 Resolution B3 (ref.~\cite{Prsa+16aj}). \label{tab:absdim}}
\begin{tabular}{lr@{\,$\pm$\,}lr@{\,$\pm$\,}l}
{\em Parameter}        & \multicolumn{2}{c}{\em Star A} & \multicolumn{2}{c}{\em Star B}    \\[3pt]
Mass ratio   $M_{\rm B}/M_{\rm A}$          & \multicolumn{4}{c}{$0.981 \pm 0.012$}         \\
Semimajor axis of relative orbit (\Rsunnom) & \multicolumn{4}{c}{$40.06 \pm 0.23$}          \\
Mass (\Msunnom)                             &  4.674  & 0.091       &  4.586  & 0.084       \\
Radius (\Rsunnom)                           &  4.117  & 0.039       &  3.653  & 0.045       \\
Surface gravity ($\log$[cgs])               &  3.879  & 0.007       &  3.974  & 0.010       \\
Density ($\!\!$\rhosun)                     &  0.0670 & 0.0015      &  0.0940 & 0.0031      \\
Synchronous rotational velocity ($\!\!$\kms)& 21.57   & 0.20        & 19.15   & 0.24        \\
Effective temperature (K)                   & 14870   & 500         & 14940   & 500         \\
Luminosity $\log(L/\Lsunnom)$               &  2.873  & 0.059       &  2.778  & 0.059       \\
$M_{\rm bol}$ (mag)                         &$-$2.44  & 0.15        &$-$2.20  & 0.15        \\
Interstellar reddening \EBV\ (mag)          & \multicolumn{4}{c}{$0.44 \pm 0.05$}			\\
Distance (pc)                               & \multicolumn{4}{c}{$1814 \pm 37$}             \\[3pt]
\end{tabular}
\end{table}



We determined the physical properties of \targ\ using the {\sc jktabsdim} code \cite{Me++05aa} and the results from the light and RV curve analyses given above. The masses are measured to 1.9\% precision and the radii to 0.9\% (star~A) and 1.2\% (star~B) precision. When comparing to LCS04 we find our results are in good agreement but with the identities of the two stars interchanged. The pseudo-synchronous rotational velocities of the stars are consistent with the values of $22 \pm 2$\kms\ and $21 \pm 2$\kms\ measured by LCS04.

The photometric analysis of LCS04 proceeded with the primary (deeper) eclipse being at phase zero; the secondary eclipse was at phase 0.62 in agreement with the current work. They then chose to swap the two stars to make the primary the hotter of the two; this also made it the smaller, less massive and less luminous component. Our analysis proceeded in the same way but without the swap, so our star~A is the larger, cooler and more massive object. The primary eclipse is deeper than the secondary eclipse, despite the inverted \Teff\ ratio, because a larger projected stellar area is eclipsed at primary than secondary. A good example of this situation can be found in our recent analysis of V454~Aur \cite{Me24obs3}.

LCS04 settled on a mean \Teff\ for the system of $14,900 \pm 500$~K from a set of calibrations based on $UBV$ and $uvby$ photometry, which is consistent with but slightly below the expected value for B5~V stars \cite{Popper80araa,PecautMamajek13apjs}. Combining this value with the ratios of the surface brightnesses and radii from Table~\ref{tab:jktebop}, and equations 5 and 6 from Southworth~\cite{Me24obs3}, gives \Teff\ values of $14,870 \pm 500$ and $14,940 \pm 500$~K for stars A and B, respectively. These values are given in Table~\ref{tab:absdim} and are much closer together than those measured by LCS04, as expected from the surface brightness ratio being only slightly above unity.

We used the results in Table~\ref{tab:absdim}, combined with the $BV$ and $JHK_s$ apparent magnitudes from Table~\ref{tab:info} and the bolometric corrections from Girardi et al.\ \cite{Girardi+02aa}, to determine the distance to \targ. The 2MASS $JHK_s$ observations were taken at orbital phase 0.268 so correspond to the out-of-eclipse brightness of the system. An interstellar reddening value of $\EBV = 0.44 \pm 0.05$~mag is needed to align the $BV$ and $JHK_s$ distances, in good agreement with the $\EBV = 0.50 \pm 0.08$~mag suggested by the {\sc stilism} reddening maps\footnote{\texttt{https://stilism.obspm.fr/}} \cite{Lallement+18aa}. The most precise distance estimate from this work is in the $K_s$ band and is $1814 \pm 37$~pc, slightly shorter than the \gaia\ DR3 \cite{Gaia21aa} value of $1948 \pm 73$~pc (a difference of 1.8$\sigma$). We are confident in our measurement of the radii of the stars -- especially in their sum, which is more important than the ratio for distance measurement -- so the discrepancy could indicate that the \Teff\ values of the stars are higher than inferred by LCS04. We experimented with adding a plausible 1000~K to the \Teff\ values, finding that this required an extra 0.01~mag of \EBV\ and gave a distance larger by 54~pc. This partial solution to the issue could be checked by obtaining a spectroscopic estimate of the \Teff\ values of the stars.



\section*{Summary and conclusions}

\targ\ is a dEB containing two B5~V stars in a orbit of period 9.653~d and eccentricity 0.192. We used light curves from seven sectors of observations using TESS, combined with spectroscopic results from LCS04, to determine the physical properties of the system. Our results are in good agreement with those of LCS04 save for an interchange of the identities of the two stars: the primary star in the current work is the larger and more massive of the two, but has evolved to be the cooler component. That the primary (deeper) eclipse corresponds to the obscuration of the cooler star is a result of the orientation of the eccentric orbit, which causes a greater stellar area to be eclipsed during primary than secondary eclipse. The precision of our results is limited by the ratio of the radii, which is poorly measured from the deep but partial eclipses produced by the system, and the scatter in the available RVs.

We find a distance to the system of $1814 \pm 37$~pc, 1.8$\sigma$ shorter than the distance of $1948 \pm 73$~pc from the \gaia DR3 parallax. A possible solution to this difference is that the stars are hotter than given in Table~\ref{tab:absdim}. The system deserves detailed spectroscopic study in order to check and confirm the \Teff\ values, measure more precise RVs to help the determination of the masses, and obtain a new spectroscopic light ratio to better determine the radii of the stars.

We compared the measured properties of \targ\ to the predictions of the {\sc parsec} theoretical stellar evolutionary models \cite{Bressan+12mn} to check the level of agreement between observation and theory, and to infer the age of the system. A metal abundance of $Z = 0.017$ and an age of $87 \pm 5$~Myr provides excellent agreement with the measured \Teff\ values and acceptable agreement with the measured radii.

The properties of both components are in the range where slowly-pulsating B-stars are found \cite{Waelkens91aa,Pamyatnykh99aca,Walczak+15aa}, promting us to conduct a search for pulsations. The data from TESS sectors 57 and 58 were chosen as they provide the longest quasi-contiguous temporal coverage, a {\sc jktebop} fit was performed, and the residuals of the fit fed to the {\sc period04} code \cite{LenzBreger05coast}. No significant periodicities were found, with 3$\sigma$ limits of 0.2~mmag for frequencies from 0.4\cd\ to the Nyquist limit (359\cd) and 1~mmag for frequencies of 0.0--0.4\cd.

A final remark is that our work has failed to significantly improve the measurements of the properties of \targ\ from the previous analysis by LCS04. The huge advance in the quality of the available light curves was not useful because the ratio of the stellar radii remains poorly determined due to degeneracies between fitted parameters. A detailed spectroscopic analysis is recommended instead, and the reader is reminded that it is good scientific practice to publish results even if they are uninteresting \cite{Rosenthal79}, especially if they act as independent confirmation of existing work \cite{Munafo+17nhb}.


\section*{Acknowledgements}

We thank the anonymous referee for a positive report.
This paper includes data collected by the TESS\ mission and obtained from the MAST data archive at the Space Telescope Science Institute (STScI). Funding for the TESS\ mission is provided by the NASA's Science Mission Directorate. STScI is operated by the Association of Universities for Research in Astronomy, Inc., under NASA contract NAS 5–26555.
This work has made use of data from the European Space Agency (ESA) mission {\it Gaia}\footnote{\texttt{https://www.cosmos.esa.int/gaia}}, processed by the {\it Gaia} Data Processing and Analysis Consortium (DPAC\footnote{\texttt{https://www.cosmos.esa.int/web/gaia/dpac/consortium}}). Funding for the DPAC has been provided by national institutions, in particular the institutions participating in the {\it Gaia} Multilateral Agreement.
The following resources were used in the course of this work: the NASA Astrophysics Data System; the SIMBAD database operated at CDS, Strasbourg, France; and the ar$\chi$iv scientific paper preprint service operated by Cornell University.



\end{document}